\documentclass[sigconf]{acmart}

\usepackage{glossaries}
\usepackage{siunitx}

\usepackage[inline]{enumitem}

\usepackage{subcaption} 

\usepackage{listings}
\usepackage{multirow}
\usepackage{xcolor} 
\usepackage{makecell}

\AtBeginDocument{%
  }

\copyrightyear{2025}
\acmYear{2025}
\setcopyright{cc}
\setcctype{by}
\acmConference[SC Workshops '25]{Workshops of the International Conference for High Performance Computing, Networking, Storage and Analysis}{November 16--21, 2025}{St Louis, MO, USA}
\acmBooktitle{Workshops of the International Conference for High Performance Computing, Networking, Storage and Analysis (SC Workshops '25), November 16--21, 2025, St Louis, MO, USA}
\acmDOI{10.1145/3731599.3767427}
\acmISBN{979-8-4007-1871-7/2025/11}

\begin{document}

\title{Seeking Cost-Optimal Infrastructure Size for Distributed Filesystems: A Ceph Case Study }

\thanks{This is the authors' accepted version of the work. The definitive Version of Record was published in SC Workshops '25, \url{https://doi.org/10.1145/3731599.3767427}.}

\author{Niccolo Tosato}
\authornote{Both authors contributed equally to this research.}
\email{niccolo.tosato@phd.units.it}
\orcid{1234-5678-9012}
\author{Isac Pasianotto}
\authornotemark[1]
\email{isac.pasianotto@phd.units.it}
\orcid{0009-0000-9759-5706}
\affiliation{%
  \institution{Univeristy of Trieste}
  \city{Trieste}
  \country{Italy}
}

\author{Ruggero Lot}
\email{ruggero.lot@areasciencepark.it}
\author{Stefano Cozzini}
\email{stefano.cozzini@areasciencepark.it}
\affiliation{%
  \institution{Area Science Park}
  \city{Trieste}
  \country{Italy}
}

\renewcommand{\shortauthors}{Tosato et al.}
\definecolor{darkpastelgreen}{rgb}{0.01, 0.75, 0.24}
\definecolor{darkpastelblue}{rgb}{0.01, 0.24, 0.75}
\newcommand{\nt}[1]{   {\color{darkpastelgreen}nt[#1]}  }
\newcommand{\ip}[1]{   {\color{darkpastelblue}ip[#1]}  }

\newcommand{\ceph}{\texttt{Ceph }}
\newcommand{\cephfs}{\texttt{CephFS }}
\newcommand{\fio}{\texttt{FIO }}

\newcommand{\IO}{I/O }
\newcommand{\iops}{IOPS }
\newcommand{\perf}{\texttt{iperf3 }}
\newcommand{\linux}{Linux }
\newcommand{\hotplug}{hotplug }
\newcommand{\hotunplug}{hotunplug }
\newcommand{\raid}{RAID }
\newcommand{\filesystem}{filesystem }
\newcommand{\filesystems}{filesystems }
\newcommand{\fss}{filesystems }

\newacronym{hpc}{HPC}{High-Performance Computing}
\newacronym{cgroups}{Cgroups}{\linux Control Groups}
\newacronym{dfs}{DFS}{Distributed Filesystem}
\newacronym{dfss}{DFS}{Distributed Filesystems}
\newacronym{pg}{PG}{Placement Group}
\newacronym{dimm}{memory modules}{DIMMs}


\sisetup{per-mode=symbol}

\DeclareSIUnit{\transfer}{T}
\DeclareSIUnit{\MTps}{\mega\transfer\per\second}
\DeclareSIUnit{\iops}{IOPS}

\glsdisablehyper

\begin{abstract}

\gls{dfss} are a crucial component of modern computing environments, and their performance is critical to the success of all the facilities that rely on them. 
However, predicting the \gls{dfss} \IO performance solely based on the storage system hardware is not trivial.

In this paper, we address this challenge by presenting an empirical method that tries to quantitatively assess how hardware configuration choices influence the performance of a \gls{dfss} using \ceph as a case study.
We investigate the influence of three hardware parameters---number of CPU cores, amount of RAM, and disk bandwidth. To control these variables, we relied on the \linux \hotplug interface and Cgroups, avoiding additional software overhead.

Our results reveal that for the analyzed workloads, decreasing hardware resources does not always yield proportional performance losses.
This method offers practical insights for designing cost-effective distributed storage systems, remaining general enough to be applied to other filesystems. 
\end{abstract}

\ccsdesc[500]{Computer systems organization~Cloud computing}
\ccsdesc[300]{Hardware~Testing with distributed and parallel systems}

\keywords{\ceph, Filesystem, \fio, Micro-Benchmark, Distributed Filesystem}


%
%

\maketitle

\glsresetall

\section{Introduction}

Storage performance is a critical aspect of \gls{hpc} systems. Once computational bottlenecks are addressed, \IO operations often become the primary limiting factor in overall system throughput and efficiency \cite{Hedges2005}. Consequently, designing an effective storage infrastructure is essential to sustain computational load and avoid performance degradation caused by \IO bottlenecks.
However, evaluating and predicting the performance of a distributed storage system is notoriously complex \cite{Tarasov2011}. 

Even more challenging is performance forecasting for a hypothetical or not yet deployed infrastructure, as theoretical models typically provide only loose upper bounds (e.g., the roofline model for \filesystems \cite{Chasapis2019}) that often diverge significantly from real-world behavior. In addition, micro-benchmarks in worst-case scenarios can show poor efficiency, as low as \qty{0.025}{\percent} \cite{Chasapis2019}, which is orders of magnitude below theoretical performance.

The intricate interplay of \IO storage components affects the performance of \gls{dfss}. 
These components include the storage media itself (HDDs, SSDs, NVMes), storage controllers (e.g., JBOD, \raid), interconnects (SATA, SAS, PCIe), the CPU, memory hierarchy, and network interface—all of which contribute to the system's \IO behavior. On the software side, performance is shaped by the operating system kernel, device drivers, \IO libraries, \gls{dfs} software, network stack, and client-side applications. \cite{Carns2020}

We propose a method to investigate the impact of three key hardware parameters:
\begin{enumerate*}[label=(\roman*)]
    \item CPU core count,
    \item available memory, and
    \item disk performance 
\end{enumerate*}
—on sequential and random \IO operations.
The choice of (starting) focusing specifically on those parameters was dictated by the fact that these three may significantly impact the hardware cost.
Our method allows us to control the host resources without introducing additional virtualization layers---i.e., VMs.
In place of emulating less powerful systems through a hypervisor, we leverage native \linux features—such as \gls{cgroups} and the CPU/RAM \hotplug interface—to constrain hardware resources directly on bare metal.
This approach removes the performance overhead typically introduced by virtualization, which can significantly impact latency and bandwidth—two critical factors in \IO intensive tasks \cite{Xavier2013}. In addition, there is no need to reprovision, redeploy, or restart the nodes.

We apply our proposed method to a \ceph \gls{dfs}  \cite{Weil2006} using \num{10} storage nodes, each one equipped with \num{12} mechanical hard drives, \num{2} NVMe drives, and InfiniBand network adapters. Using up to \num{8} compute nodes as clients, we perform extensive benchmarking with \fio \cite{Axboe2016}.

\section{Background}
\subsection{\ceph}

\ceph \cite{Weil2006} is a distributed storage platform that provides access to storage in multiple forms: object storage via S3-compatible APIs, block storage (RBD), and a POSIX-compliant filesystem: \cephfs. 
Its flexibility makes it suitable for a range of environments—from cloud providers to hyperscalers and \gls{hpc} facilities. \ceph allows storage capacity and performance to scale horizontally by adding more storage nodes.
Files stored in \cephfs are internally split into objects. Each object is assigned to a \gls{pg}, which belongs to a pool. The pool defines rules for \gls{pg} assignment to storage devices, including the count of replicas, replica placement strategy, and device type (e.g., HDDs or NVMes).

\ceph ensures data redundancy using either replication or erasure coding. In this study, we focus on replicated storage. Each replica is placed on a different node to reflect a production-grade environment designed to tolerate node failures and prevent data loss.

\subsection{Filesystem and performance measurement}

Modeling, measuring, and forecasting \filesystem performance remains a notoriously difficult challenge due to the intricate interactions between software layers, network behavior \cite{Carns2020}, and underlying storage hardware \cite{Traeger2008, Tarasov2011}.

In addition to empirical evaluations, several theoretical models have been proposed to predict performance under specific storage coding schemes (e.g., erasure coding \cite{Li2016}) or general \gls{dfs} abstractions \cite{Wu2014}, eventually using roofline-inspired frameworks to characterize system limits given a specific hardware \cite{Chasapis2019}.
Others have empirically examined the comparative performance of different \fss \cite{Lee2021} using standard workload patterns (e.g., random vs. sequential \IO, varying block sizes, differing numbers of clients) \cite{Gudu2014, Zhang2019}.

The systems community has gradually established a set of commonly considered best practices for \IO benchmarking, which emphasize selecting representative workloads, setting the focus on what has to be measured (e.g., measure actual throughput, avoiding caches), ensuring experimental reproducibility, and result reliability. \cite{Tarasov2011, Traeger2008}.
Moreover, synthetic workloads, while helpful in isolating specific effects, rarely capture the diversity of real-world access patterns \cite{Carns2020}, and strong benchmark results do not necessarily translate to optimal performance for actual applications, which may favor alternative storage architectures  \cite{Li2016}.

Furthermore, performance is influenced by a wide array of factors beyond \filesystem configuration or hardware specifications.
Among these factors we can mention the “aging” effect of the \filesystem \cite{Smith1997}, the level of client concurrency \cite{Wu2014} (modeled) \cite{Gudu2014} (empirical), the workload composition (e.g., read/write ratio) \cite{Paridon2010}, and application-level optimizations—all of which can significantly affect throughput and latency.
Other studies have focused on fine-grained \filesystem parameters—such as journaling strategies, caching mechanisms, and striping configurations—and their impact on performance, particularly in systems like CephFS \cite{Bhat2024, Zhang2019, Gudu2014}.

To the best of our knowledge, no prior work provides a comprehensive empirical evaluation of how the entire hardware stack jointly shapes \gls{dfs} performance. This paper aims to address that gap by systematically quantifying how CPU, memory, and storage devices together define the performance envelope of \gls{dfs}.

\section{Methods}

First, we assessed the single-machine performance, ensuring that all components involved in \IO operations behave correctly (\autoref{subsec:single-node}). This experiment allows us to identify potential bottlenecks in the infrastructure \cite{Carns2020} and enables us to calculate the maximum theoretical \IO bandwidth delivered, neglecting software limitations (\autoref{subsec:theoretical}), by a single node  \cite{Chasapis2019}.
After that, we evaluated the performance of a multi-node setup by deploying \ceph \filesystem and using a remote \fio client to measure bandwidth and \iops while continuously reducing the number of cores, RAM, and  hard drive speed; the effects of these parameters are theoretically explored in \cite{Wu2014}.

\begin{table}[htbp]
\centering
\caption{Technical specification of compute and storage nodes. Compute nodes are used to run \fio jobs that access the \cephfs file system provided by the storage nodes. To avoid metadata bottlenecks, two dedicated compute nodes are reserved for hosting the \ceph Metadata Server (MDS).}
\label{tab:hardware}
\begin{tabular}{lc}
\toprule
\textbf{Component} & \textbf{Compute Node} \\
\midrule
Model &  Dell PowerEdge R6625  \\
CPU &  \num{2}$\times$ AMD EPYC 9374F 32C \\
Memory & \num{16}$\times$\qty{32}{\giga\byte} DDR5 @ \qty{4800}{\MTps} \\
Infiniband & Mellanox ConnectX-7 @ \qty{200}{\giga\bit\per\second} \\
\\
\toprule
\textbf{Component} & \textbf{Storage Node} \\
\midrule
Model &  Dell PowerEdge R760 \\
CPU & \num{2}$\times$ Intel Xeon Gold 6426Y 16C \\
Memory & \num{12}$\times$\qty{16}{\giga\byte} DDR5 @ \qty{4800}{\MTps} \\
Infiniband & Mellanox ConnectX-7 @ \qty{200}{\giga\bit\per\second} \\
\midrule
Hard drive & \num{12}$\times$\qty{22}{\tera\byte} Western Digital \\ & WUH722222AL5200 \\
NVMe & \num{2}$\times$\qty{15.36}{\tera\byte} Dell \\ & ISE PS1010 RI U.2  \\
Disk controller & Dell HBA355i \\

\bottomrule
\end{tabular}
\end{table}

\subsection{Single-node evaluation}
\label{subsec:single-node}
We assessed the performance of a single storage node, equipped with the hardware detailed in \autoref{tab:hardware}, focusing on the
\begin{enumerate*}[label=(\roman*)]
    \item storage media,
    \item storage adapter and its interconnect, and
    \item network interface.
\end{enumerate*}
By evaluating these components, we can determine an upper bound for theoretical performance based on the given hardware, and by doing so, we can find out which component will be the bottleneck at the single-server level. 

\paragraph{Storage Device}

We used \fio \cite{Axboe2016} as a micro-benchmark tool; we set a block size of \qty{4}{\mebi\byte} and \qty{4}{\kibi\byte}, respectively, for sequential and random workloads. These block size values were selected to align with those reported in the devices' datasheet. Each measure is repeated \num{5} times to collect enough statistics. 
The results are summarized in \autoref{tab:hdd}, where we report the minimum, average, and maximum values achieved among the different runs.
We tested both raw \IO and \texttt{XFS}, as it is the most performant \filesystem commonly used in production for evaluating a single device.
We observe that the maximum sequential read and write performance (respectively \qty{260.07}{\mebi\byte\per\second} and \qty{258.10}{\mebi\byte\per\second} using \texttt{XFS} filesystem) is consistent with the drive specifications, which are rated at \qty{277}{\mebi\byte\per\second} for sustained transfer.

\begin{table}[h]
  \caption{Measured performance of mechanical hard drives in terms of sequential \IO bandwidth (block size: \qty{4}{\mebi\byte}) and random \iops (block size: \qty{4}{\kibi\byte}). Tests were performed both on raw devices and using the \texttt{XFS} filesystem. Each test was repeated \num{5} times. For each test the minimum, average, and maximum values are reported. The slightly higher performance observed with \texttt{XFS} may be due to the filesystem’s ability to manage \IO requests more efficiently (e.g, by reordering and merging them, even under direct \IO).
  }
  \label{tab:hdd}
  \begin{tabular}{llcc}
    \toprule
     &   & \makecell{\textbf{Sequential \IO} \\  \textit{min}/\textit{avg}/\textit{max}  \lbrack MiB/s\rbrack} & \makecell{\textbf{Random \IO} \\ \textit{min}/\textit{avg}/\textit{max} \lbrack IOPS\rbrack } \\
    \midrule
    \multirow{2}{*}{\makecell{Raw}} 
      & Read & \num{252}/\num{252}/\num{253} & \num{169}/\num{414}/\num{701} \\
      & Write & \num{223}/\num{250}/\num{252} & \num{1036}/\num{1098}/\num{1184} \\
    \midrule
    \multirow{2}{*}{\texttt{XFS}} 
      & Read & \num{258}/\num{260}/\num{260} & \num{168}/\num{425}/\num{708} \\
      & Write & \num{220}/\num{255}/\num{258} & \num{1036}/\num{1115}/\num{1192} \\
    \bottomrule
  \end{tabular}
\end{table}

\paragraph{Storage Controller}

Then we verified that each storage device can handle \IO operations continuously, even when devices operate in parallel, and that their aggregate performance scales with the number of devices to ensure that the storage controller \emph{Dell HBA355i} and its interconnect are not limiting factors for device access. 
 We used \fio to generate sequential \IO workloads, with multiple processes accessing different drives concurrently. We tested both write and read operations using a block size of \qty{8}{\kibi\byte}. Each \fio job writes an \qty{8}{\gibi\byte} file, and each test operation runs for at least \num{60} seconds; runs are repeated \num{15} times\footnote{The number of run repetitions in our experiments varies (\num{5}, \num{7}, and \num{15}) depending on the execution time of each benchmark. To minimize benchmarking duration and return the hardware to production as quickly as possible, while still ensuring meaningful analysis, we adjusted the number of runs to balance statistical reliability with practical feasibility.}.

\begin{figure}[h]
  \centering
  \includegraphics[width=\linewidth]{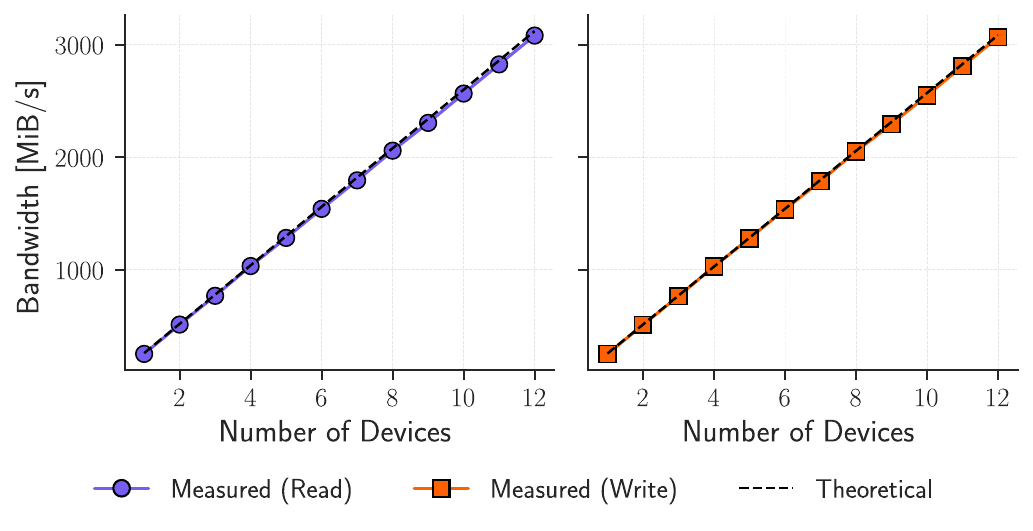}
  \caption{Aggregated bandwidth measured while scaling up to \num{12} mechanical drives. The measured performance closely aligns with the theoretical values, confirming that the disk controller can handle concurrent access to all devices without introducing performance degradation.}
    \label{fig:hba}
\end{figure}

The \emph{Dell HBA355i} controller achieves a maximum bandwidth of \qty{3082}{\mebi\byte\per\second} and \qty{3075}{\mebi\byte\per\second} respectively during sequential read and write. 
Using the maximum value reported in \autoref{tab:hdd}, we can then estimate the maximum theoretical bandwidth. The ratio between the observed value and the theoretical value is \num{0.92} for a sustained transfer rate with \num{12} HDDs.
The \emph{scalability} is almost perfect up to \num{12}, since theoretical bandwidth and the measured one are almost overlapping each other, as reported in \autoref{fig:hba}. 

\paragraph{Network Interface}
Since all communication between storage nodes and clients occurs over the network, also this aspect must be taken into account; hence, we analyzed the performance of the network interface.
The expectation is that the network's maximum bandwidth exceeds the aggregate bandwidth of the storage drives; otherwise, the theoretical performance of the \filesystem is bounded by communication\cite{Chasapis2019}.
Since we used an Infiniband adapter, we tested first bare RDMA performance, to assess the IB transport layer using \texttt{perftest}'s benchmarks.

Benchmark results for point-to-point communication using RDMA, as shown in \autoref{tab:infiniband}, indicate a best-case write latency of \qty{1.3}{\micro\second} and a read bandwidth of \qty{176.1}{\giga\bit\per\second}. 
The values obtained are consistent with the vendor's declared performance specifications (i.e., \qty{200}{\giga\bit\per\second} as a bandwidth reference). The latency is not disclosed for the \textsc{NVIDIA Quantum-2 QM9700} switch used in the test; however, latencies on the order of \unit{\micro\second} are considered satisfactory for NDR Infiniband communications.


Next, we evaluated TCP performance over InfiniBand using the \texttt{IPoIB} driver and interface, since client and nodes communicate using TCP protocol. In this test, the negotiated \texttt{IPoIB} link speed is \qty{100}{\giga\bit\per\second}, despite the interface being rated at \qty{200}{\giga\bit\per\second} \footnote{This is due to a technical limitation: the Subnet Manager is running on a device that does not support NDR-mode InfiniBand, limiting \texttt{IPoIB} negotiation speed.}. We used \perf with \num{4} parallel streams, achieving a total bandwidth of \qty{92.7}{\giga\bit\per\second}.

\begin{table}[ht]
\centering
\caption{Infiniband latency and bandwidth measurements obtained using the \texttt{perftest} tool to evaluate RDMA performance.
Results represent the average over at least \num{5000} iterations, performed using the Reliable Connection (RC) transport mode.}
\label{tab:infiniband}
\begin{tabular}{lcc}
\toprule
  & \textbf{Latency [\unit{\micro\second}]} & \textbf{Bandwidth [\unit{\giga\bit\per\second}]} \\
\midrule
IB Write           & \num{1.29}            & \num{175.39}                       \\
IB send            & \num{1.45}            & \num{175.65}                       \\
IB read            & \num{2.60}            & \num{176.09}                       \\
\bottomrule
\end{tabular}
\end{table}

\subsection{Theoretical performance and efficiency}
\label{subsec:theoretical}

The single-node performance analysis presented in \autoref{subsec:single-node} provides a baseline for estimating the theoretical performance of the 10-node distributed file system used in this paper. Considering the aggregated theoretical speed of the hard drives (\qty{3324}{\mebi\byte\per\second}), the measured storage controller throughput of approximately \qty{3100}{\mebi\byte\per\second} (i.e., an efficiency above 92\%), and a network interface rated at \qty{12.5}{\giga\byte\per\second} we conclude that the aggregate device bandwidth effectively impact the theoretical single-node performance. 
Assuming linear scalability with respect to the number of nodes, the theoretical peak performance of the \gls{dfs} can be estimated as:

\noindent ~$\text{single node bandwidth} \times \text{\# of nodes} \simeq\ $~\qty{32.5}{\gibi\byte\per\second}.

Then, we established a performance baseline of the \ceph \filesystem for replication factors \num{1}, \num{2}, and \num{3} using the storage node configuration described in \autoref{tab:hardware}, with no constraints on RAM, CPU cores, or drive speed.
By comparing the measured sequential read and write performance against the theoretical peak performance, we compute the efficiency of the system, as reported in \autoref{tab:efficiency}. We can note that with no replicate write the performance that we achieve is \qty{48}{\percent} of the maximum peak performance, highlighting a significative performance gap.

\begin{table}[h]
\caption{Efficiency of the \ceph \gls{dfs}, calculated as the ratio between the sequential bandwidth achieved using a \num{10} nodes \ceph cluster (without resource constraints) and the theoretical peak performance of \qty{32.5}{\gibi\byte\per\second}, as defined in \autoref{subsec:theoretical}. }
\label{tab:efficiency}

\begin{tabular}{ccc}
\toprule
  Replication factor & Read efficiency & Write efficiency \\
\midrule
\num{1}  & \num{0.144} & \num{0.479} \\
\num{2}  & \num{0.138} & \num{0.258} \\
\num{3}  & \num{0.141} & \num{0.176} \\
\bottomrule
\end{tabular}
\end{table}

\subsection{Limiting \IO using \gls{cgroups}}

To control \IO bandwidth, we leveraged \gls{cgroups} \cite{Kerrisk20210}, which allow for monitoring and limiting resource usage by processes. This decision was inspired by the fact that \gls{cgroups} are widely adopted by container orchestrators (e.g., Kubernetes), runtimes to manage resource access in containerized environments, and job schedulers (e.g., Slurm).

\gls{cgroups} support multiple \emph{controllers} (also called \emph{subsystems}) to manage various hardware resources. In our work, we used the \IO controller (known as \emph{blkio} and as \emph{io} controller in \gls{cgroups} v2). 
The \emph{blkio} controller exposes several tunable parameters. We made use of the throttling policy via two specific files:  ~\path{blkio.throttle.read_bps_device} and ~\path{blkio.throttle.write_bps_device} exposed in the \emph{cgroupfs} pseudo-\filesystem.

This interface allowed us to set bandwidth limits during read and write operations for a specific process and block device, example of interface usage is reported in \autoref{lst:cgroup}.

\begin{lstlisting}[caption={Example showing how to limit device read and write throughput in bytes per second. Devices are identified using the \linux kernel's block device naming convention, represented as a <major>:<minor> number pair.}, label={lst:cgroup}]
$ echo "<major>:<minor> <rate_bytes_per_second>" > /cgrp/blkio.throttle.write_bps_device
$ echo "<major>:<minor> <rate_bytes_per_second>" > /cgrp/blkio.throttle.read_bps_device
\end{lstlisting}

Starting with version $15$ (\textsc{Octopus}), the \ceph file system offers full support for containerized deployments, with each \ceph daemon running inside a dedicated container managed by Podman. The components responsible for performing \IO operations on physical storage devices are known as Object Storage Daemons (OSDs), with each OSD managing a single device. We utilized Podman's \gls{cgroups} options to throttle the bandwidth available to each OSD, thereby controlling the \IO throughput of the underlying physical devices.
 We applied \gls{cgroups}-based bandwidth limitations towards hard drives to all corresponding containers, an example using Podman is provided in \autoref{lst:podman}.

\begin{lstlisting}[caption={Updating \ceph container resource limits using Podman. Podman leverages \linux \gls{cgroups} to constrain container resource allocation. In this snippet, we apply updated \gls{cgroups} limits to the running \ceph container.}, label={lst:podman}]
$ podman update <container-id>  --device-write-bps /dev/sda:<speed>
\end{lstlisting}

Note that no bandwidth limitations were applied to the NVMe devices, as they are used exclusively for metadata storage and their performance impact is beyond the scope of this work.

\subsection{Offlining resources via \hotplug interface}

The \linux kernel implements \hotplug interfaces for both CPUs and memory. These interfaces are typically used to disable or enable physical CPU cores and \gls{dimm} at runtime—for maintenance or power-saving purposes—but in this work, we leveraged them to simulate lower-end hardware configurations.
 
\paragraph{CPU \hotplug}
We used the CPU \hotplug interface to simulate systems with fewer available cores without relying on virtualization, and this enables us to test different CPU topologies (e.g., power off one entire socket or halve the number of cores for each socket). 
The interface was controlled through \texttt{sysfs}, allowing individual cores to be toggled online or offline, an example is provided in \autoref{lst:cpu_hotplug}

\begin{lstlisting}[caption={Example of turning a specific core \texttt{<core\_id>} on and off using the \linux \hotplug interface. Changes take effect immediately. }, label={lst:cpu_hotplug}]
$ echo "0" > /sys/devices/system/cpu/cpu<core_id>/online 
$ echo "1" > /sys/devices/system/cpu/cpu<core_id>/online 
\end{lstlisting}

Note that depending on the system's hardware, disabling cores may affect core frequency scaling and power management behavior. Reducing the number of active cores could reduce the maximum Thermal Design Power (TDP) and consequently the maximum core frequency increase, favouring single-thread component performance.

\paragraph{Memory \hotplug} 
The memory \hotplug feature allows for dynamic resizing of the physical memory available to the system at runtime. It is primarily used for replacing faulty \gls{dimm} or optimizing energy usage by disabling memory banks.

The \linux kernel exposes a \emph{sysfs} interface to offline memory blocks (the smallest unit of memory that can be disabled). We use this \hotplug mechanism to reduce available memory and simulate systems with smaller memory. An example demonstrating the use of this interface is shown in \autoref{lst:mem_hotplug}.

\begin{lstlisting}[caption={Example of memory \hotunplug: offlining a specific memory block <block\_id>. Page migration is involved; if page-locked memory is allocated in this block, the offlining procedure will fail.}, label={lst:mem_hotplug}]
$ echo "0" > /sys/devices/system/memory/memory<block_id>/online
\end{lstlisting}

As with CPU \hotplug, this technique allowed us to simulate a variety of memory layouts (e.g., reducing the memory evenly per memory channel or diminishing the amount of populated channels), which can have significant effects on performance.

\subsection{Multi-node benchmark}

We deployed a  \ceph \num{18} (\textsc{Reef}) cluster using \num{10} storage nodes (configuration detail in \autoref{tab:hardware}), achieving a maximum \filesystem raw size of \qty{2.640}{\peta\byte}. We created two replicated pools: one enforcing \num{2} replicas and the other with \num{3} replicas; both placing replicas on different hosts and storing data on HDDs. As metadata performance is not the focus of this work, we ensured that metadata servers (MDS) were not a bottleneck by deploying two dedicated MDS nodes that are not constrained in terms of resources and performance. 
For the same reason, we have chosen to store that metadata on a separate pool backed by NVMe devices. 

To reduce interference while measuring performance, we disabled background \ceph operations like scrub and deep-scrub during benchmarking.
Storage nodes have been set up without hyper-threading, the disk controller is instructed to act just as a plain JBOD (\emph{just a bunch of disks}), so no \raid is present, and the operative system \IO scheduler used was \texttt{mq-deadline} for mechanical hard drives.

\IO benchmarking was performed using \fio \cite{Axboe2016}, running in client-server mode. All compute nodes (hardware details reported in 
\autoref{tab:hardware}) served as \fio clients. Each client performed \IO operations under the control of the \fio server, which collects and reports aggregate bandwidth and \iops metrics from \fio clients. 

\paragraph{Memory size and core count}

This experiment explores \num{16} different storage node configurations. Specifically, we evaluated nodes equipped with \numlist{32;64;128;192}\unit{\giga\byte} of RAM, combined with CPU core counts of \numlist{8;16;24;32}. Each node includes 12 HDDs and 2 NVMe drives (\autoref{tab:hardware}), resulting in \numlist{0.6;1.1;1.7;2.3} CPU cores and \numlist{2.3;4.6;9.1;13.7}\unit{\giga\byte} of RAM per OSD, depending on the configuration.
We measured \iops and bandwidth while varying the amount of available memory and the number of cores allocated per OSD. While reducing memory, we maintained a constant ratio across NUMA nodes. A similar strategy was applied to core offlining, evenly distributing active cores across CPU sockets.

We conducted each experiment on \num{8} compute nodes, with each node running 64 \fio clients (one process per core). These clients collectively targeted \num{10} storage nodes. The workload consisted of read and write operations on \num{512} files—one per \fio process—each sized at \qty{16}{\giga\byte}. Every test was repeated \num{7} times and configured to run for at least \num{150} seconds to ensure stable and consistent measurements. Due to time and infrastructure limitations, we did not collect additional statistics.

We used the \emph{posixaio} backend in \fio and took care to minimize caching effects. This was achieved by disabling caches and buffers, enforcing direct \IO, and using files significantly larger than the available system cache. An example is provided in \autoref{sec:fio-job}.

\paragraph{Hard drive speed}

In this final experiment, we evaluated \iops and bandwidth while throttling storage device throughput using \gls{cgroups} controllers via Podman. The throttling was applied in increments of \qty{25}{\mega\byte\per\second}, starting from \qty{100}{\mega\byte\per\second} and reaching up to a maximum of \qty{275}{\mega\byte\per\second}. 
To avoid impacting metadata performance, NVMe devices were left unthrottled. Additionally, no CPU or memory constraints were applied to the storage nodes. We perform this experiment across 10 storage nodes and 6 client nodes.

We used the \emph{libaio} backend in \fio to generate the \IO workload, accessing a total of \num{384} files\footnote{Unlike the previous experiment, fewer processes-and consequently fewer files-were used due to the limited availability of compute nodes (i.e., \num{6} instead of \num{8}).}, each \qty{16}{\giga\byte} in size. Cache effects were again, minimized by bypassing system caches whenever possible, and each of 7 run lasted for at least \num{150} seconds to ensure stable and sustained performance measurements.

\section{Results}

We present in this section the result on \num{10}-nodes \ceph \filesystem, measuring first the effect of core count and memory size (\autoref{subsed:memory_core_result}), then the effect of the hard drive speed (\autoref{subsed:hdd_result}).

\subsection{Ceph - Memory size and core count effect}

\label{subsed:memory_core_result}

\paragraph{Sequential \IO} 

\begin{figure}[h]
  \centering
  \includegraphics[width=\linewidth]{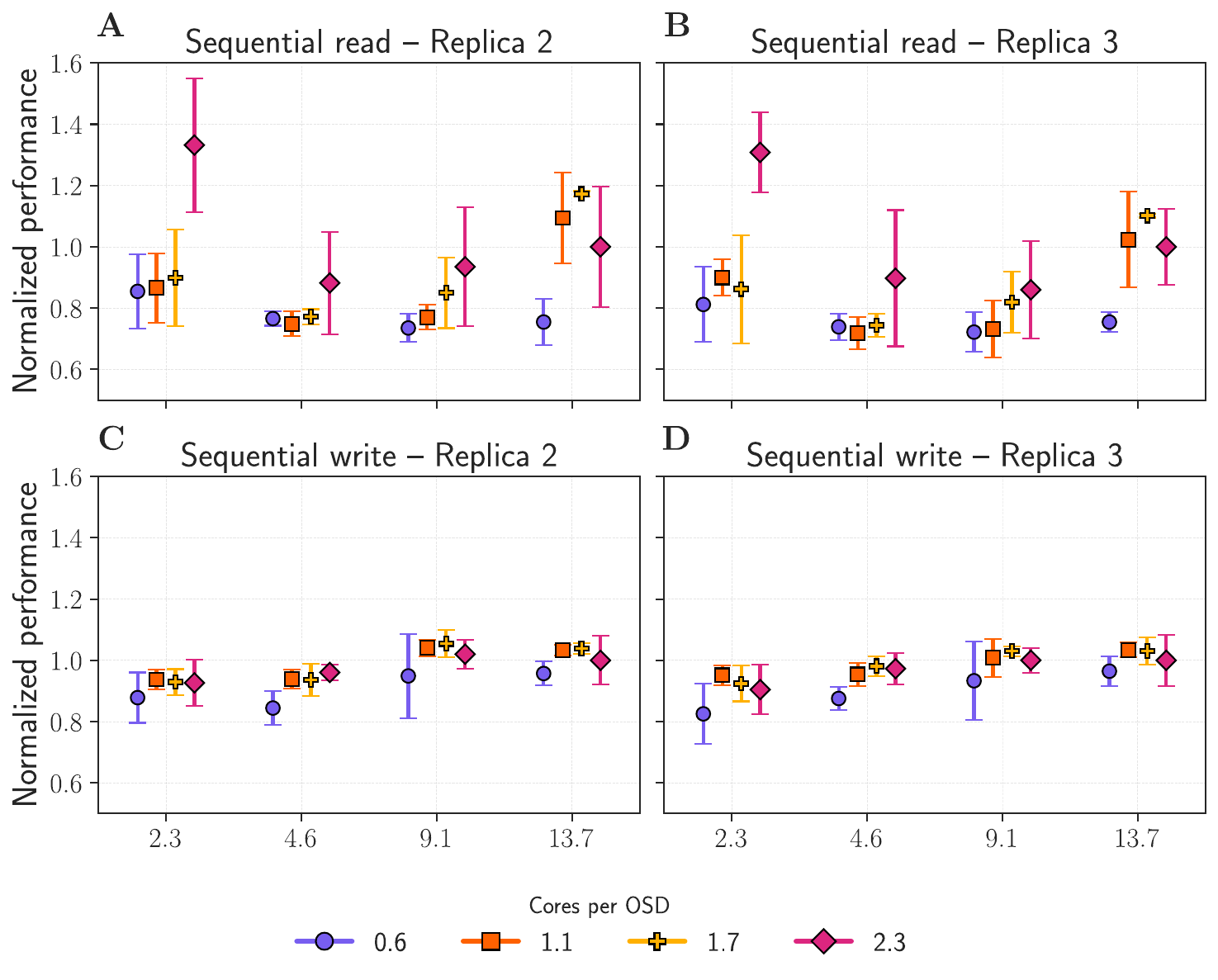}
  \caption{Normalized performance of the \cephfs \filesystem under sequential \IO workloads across different core and RAM, with replication factors of \num{2} and \num{3}. Each panel is normalized with respect to the performance baseline without hardware constraints (i.e., \num{32} cores and \qty{192}{\giga\byte} of RAM), represented by the rightmost point.
  \newline \textsc{Panel A,B}: an anomalous performance increase of approximately \qty{30}{\percent} is observed for the \qty{2.3}{\giga\byte} RAM and \num{2.3} core per OSD configuration. This is considered a potential outlier, possibly due to errors in the experimental setup or an artifact in measures (discussed in \autoref{subsed:memory_core_result}). 
    \newline \textsc{Panel C,D}: the highest performance loss is always below \qty{20}{\percent} with respect to the baseline.
  Noteworthy, the top performance is achieved often with \num{1.7} cores per OSD and not \num{2.3} as could be expected, this particularly evident in \textsc{Panel A}.
  \newline 
  Absolute values are provided in \autoref{tab:rw_details_sequential}.}
    \label{fig:sequential}
\end{figure}

The result reported in \autoref{fig:sequential} shows the normalized bandwidth achieved by running sequential IO with a block size of \qty{8}{\mebi\byte}. 
The results are normalized with respect to the measure obtained using the maximal configuration with no hardware constraints, i.e., \num{32} cores and \qty{192}{\giga\byte} of RAM.
The measurements exhibit variability, particularly high under constrained resource configurations (e.g., \num{0.6} cores per OSD or low memory per OSD) and read operations.
Interestingly, the highest core allocation (e.g., \num{2.3} cores per OSD) does not consistently yield the best performance as one could expect; moreover, it exhibits the highest noise during read tests. 

The configuration with \num{2.3} cores per OSD  and the minimum amount of RAM yields the highest read performance; however, its measurements become inconsistent under successive RAM increases and appear as outliers (likely an undetected \fio{} error).
The bandwidth reported by \fio{} is calculated by aggregating the written bytes and taking the maximum runtime over all processes. This value is inconsistently higher than the bandwidth averaged over individual \fio{} processes obtained by sampling \IO speed during the test execution\footnote{We are referring to the field \texttt{bw\_mean} reported in the JSON output of the fio server.}. We also observed that each job lasted on average more than 160 seconds (with a minimum required duration of 150 seconds), whereas each process that appeared as an outlier terminated on average after 154 seconds. This discrepancy suggests that an error not captured in the logs may have produced the observed artifact (e.g., speculating that clients crashed and returned earlier, reducing the runtime and increasing the bandwidth). For this reason, we will not take these results into account during the discussion (\autoref{fig:sequential}, Panel \textbf{A, B}).
For sequential writes—and especially for reads—the performance remains nearly constant when only half a core per OSD is used, regardless of increasing RAM. In contrast, configurations with higher core counts show consistently a noticeable performance improvement when the memory per OSD exceeds \qty{9.1}{\giga\byte}.
The results executed with a replica count of \num{2} and \num{3} are coherent and exhibit the same behavior. 

\paragraph{Random \IO}

\begin{figure}[h]
  \centering
  \includegraphics[width=\linewidth]{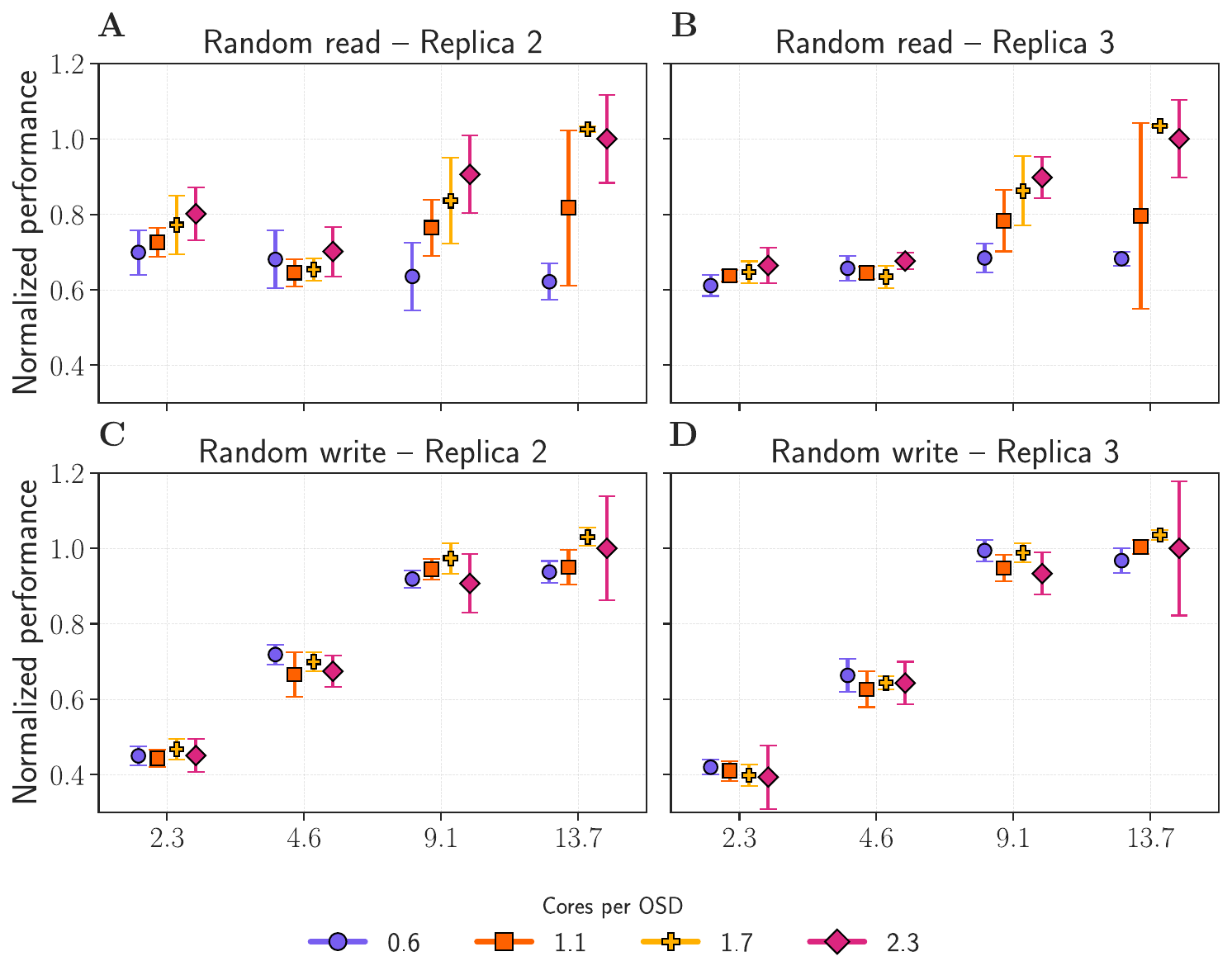}
  \caption{Normalized performance of the \cephfs \filesystem under random \IO workloads across different core and RAM configurations, with replication factors of \num{2} and \num{3}. Each subplot is normalized with respect to the performance baseline without hardware constraints(i.e., \num{32} cores and \qty{192}{\giga\byte} of RAM), represented by the rightmost point).
    \newline  \textsc{Panel A,B}: random read operation shows that decreasing the amount of cores per OSD could reduce the performance by \qty{40}{\percent}. Notably, increasing the core count is beneficial only when accompanied by a corresponding increase in RAM, highlighting a CPU-limited pattern. A consistent performance penalty of approximately \qty{35}{\percent} is observed when only \num{0.6} cores are allocated per OSD.
    \newline  \textsc{Panel C,D}: random write performance shows significant degradation when RAM is reduced. In particular, decreasing the memory to \qty{2.3}{\giga\byte} per OSD results in up to a \qty{60}{\percent} performance loss compared to the baseline. 
    \newline
    Absolute values are provided in \autoref{tab:rw_details_random}.
    }
        \label{fig:random}
\end{figure}

The result reported in \autoref{fig:random}  are obtained by performing random writes with \qty{4}{\kibi\byte} chunks, following the standard \iops definition, reported normalized with respect to the maximal configuration with \num{32} cores and \qty{192}{\giga\byte} of RAM. 
Unlike the sequential read experiment, performance variability is more evident when each OSD is allocated a larger share of system resources—especially during read operations (\autoref{fig:random}, Panel \textbf{A,B}).

Write performance (\autoref{fig:random}, Panel \textbf{C,D}) appears more stable overall, though the configuration with \num{2.3} cores per OSD still shows the highest variance as in the previous experiment and does not always show the best performance.
In contrast to the sequential workload, a clearer relationship between allocated resources and performance is observable here, particularly for random write operations.
Decreasing the amount of RAM always brings a reduction in random write performance, regardless of the core count (\autoref{fig:random}, Panel \textbf{C,D}). This does not hold for random read, since the drop in performance due to the decrement of RAM is proportional to the core count. 
Different replica counts show qualitatively the same behavior, but quantitatively increasing resources yield a better improvement in the case of higher replica counts. 

\subsection{\ceph - Hard drive speed effect}
\label{subsed:hdd_result}

In this section, we report the results obtained by varying the hard drive speed and measuring performance under both random and sequential workload, using respectively a blocksize of \qty{4}{\kibi\byte} and \qty{8}{\mebi\byte}. 

\paragraph{Sequential \IO}

\begin{figure}[h]
  \centering
  \includegraphics[width=\linewidth]{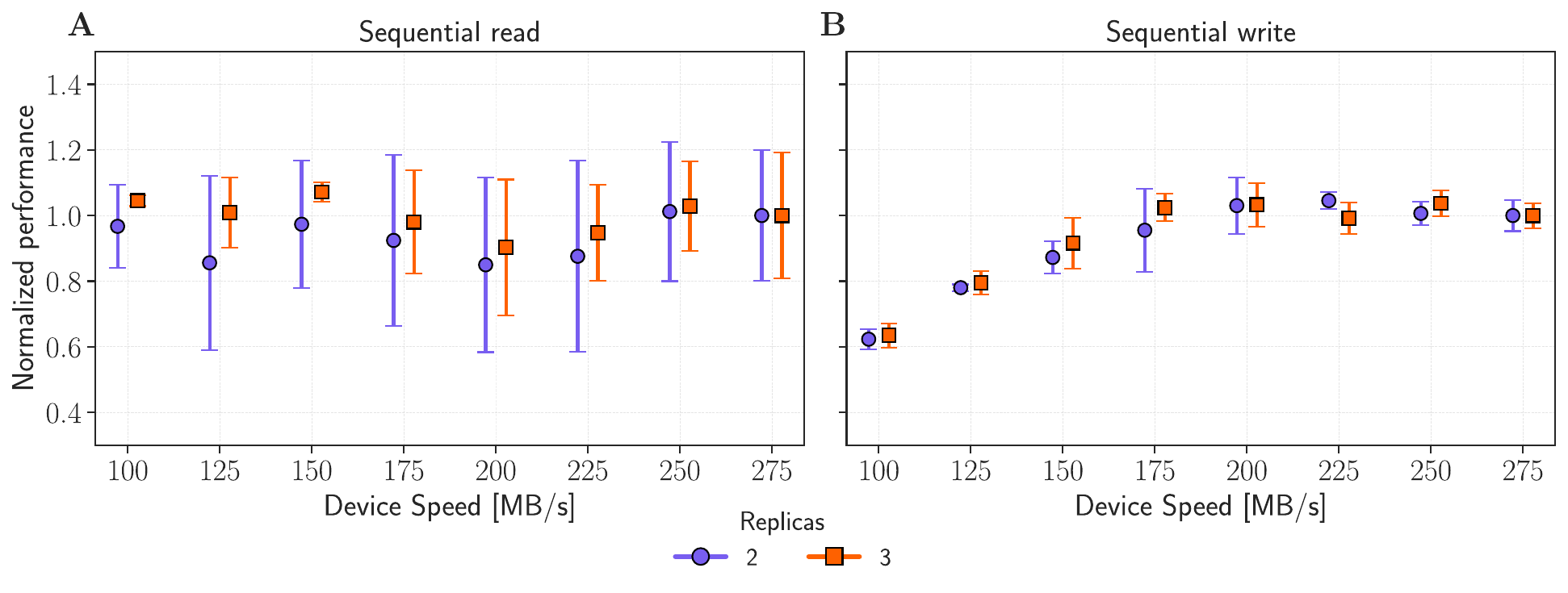}
  \caption{
   Normalized performance of the \cephfs \filesystem under sequential I/O workloads across different device speed.
  Each panel is normalized with respect to the performance obtained with the fastest disk speed available (i.e., \qty{275}{\mega\byte\per\second}).
  \newline \textsc{Panel A:} sequential read exhibits no performance drop while using slower devices; 
  \newline 
  \textsc{Panel B:} sequential write exhibits a performance plateau when device bandwidth is larger than \qty{200}{\mega\byte\per\second}. The configuration with the device speed of \qty{100}{\mega\byte\per\second} reveals a performance degradation of up to \qty{65}{\percent}.
  \newline Absolute value are provided in table \autoref{subtab:disk_det_seq_rep2} and \autoref{subtab:disk_det_seq_rep3}.
  }
\label{fig:sequential-hdd}

\end{figure}

The reading (\autoref{fig:sequential-hdd}, Panel \textbf{A}) test exhibits a consistent noise level across all levels of device speed, with the same pattern for replica 2 and replica 3 configurations. The performance is constant regardless of the device speed.

During the writing test  (\autoref{fig:sequential-hdd}, Panel \textbf{B}), we observe a low variance in bandwidth and two distinct regimes: the first regime shows a linear relationship between \filesystem bandwidth and hard drives speed. In contrast, the second regime reaches a performance plateau starting at approximately \qty{200}{\mega\byte\per\second}.

\paragraph{Random \IO}

\begin{figure}[h]
  \centering
  \includegraphics[width=\linewidth]{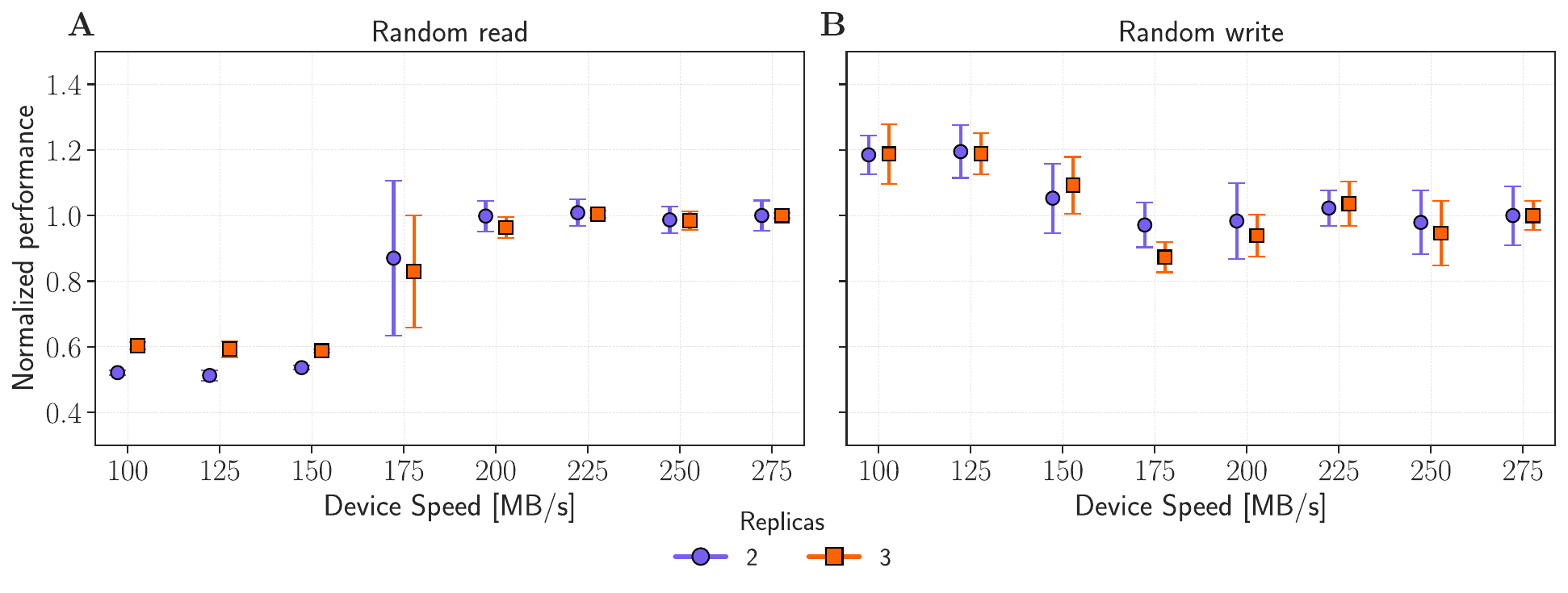}
  \caption{
     Normalized performance of the \cephfs \filesystem under random I/O workloads across different device speed.
   Each panel is normalized with respect to the performance obtained with the fastest disk speed available (i.e., \qty{275}{\mega\byte\per\second}).
  \newline \textsc{Panel A:} 
  for devices with random read speeds below \qty{150}{\mebi\byte\per\second}, performance reveals a first plateau with a \qty{40}{\percent} loss relative to the baseline. For devices faster than \qty{150}{\mega\byte\per\second}, a second plateau is reached.
\newline \textsc{Panel B:} random write exhibits no performance loss while using slower devices.
\newline Absolute values are provided in table \autoref{subtab:disk_det_seq_rep2} and \autoref{subtab:disk_det_seq_rep3}.
  }
  \label{fig:random-hdd}
\end{figure}

The read results (\autoref{fig:random-hdd}, Panel \textbf{A}) show two plateaus separated by a transient phase with high noise. The first plateau is stable, followed by a noisy transient at \qty{175}{\mega\byte\per\second}, which then reaches the second plateau at \qty{200}{\mega\byte\per\second}—matching the threshold observed in \autoref{fig:sequential-hdd}, Panel \textbf{B} during sequential writes. The first plateau exhibits half the performance of the second one. 
The write result (\autoref{fig:random-hdd}, Panel \textbf{B}) shows that performance is constant regardless of the device speed, in the same way as sequential read.

\section{Discussion}

In this discussion, we analyze the results obtained by running sequential and random I/O workloads on a \num{10}-node \ceph cluster configured with \num{16} different combinations of RAM and CPU core counts (\autoref{subsed:memory_core_result}). We then examine how variations in hard drive speed affect overall \ceph{} cluster performance (\autoref{subsed:hdd_result}); providing insights into the potential impact of specific hardware components on \filesystem performance. 

Each result is normalized against a performance baseline obtained from tests run without any resource constraints, the same baseline used to compute \gls{dfs} efficiency in \autoref{subsec:theoretical}.

\paragraph{Impact of memory size and core count}
While performing sequential \IO  the configuration with \numlist{1.1;0.6} CPU cores per OSD and the lowest amount of RAM shows a performance drop of less than \qty{20}{\percent} compared to the baseline. Notably, most of this loss occurs when the memory per OSD is reduced from \qty{13.7}{\giga\byte} to \qty{9.1}{\giga\byte}. 
Allocating only \num{0.6} CPU cores per OSD clearly reveals a CPU-bound behavior in sequential \IO, as increasing memory alone results in nearly constant performance.
The smallest performance drop is observed in sequential write workloads; the loss consistently remains well below \qty{20}{\percent} with respect to the baseline performance, even with \num{6}$\times$less RAM and \num{4}$\times$ fewer CPU cores compared to the baseline configuration. 
Interestingly, the maximum performance is not always achieved with the highest number of cores. This could be due to the lack of core affinity, causing processes to be switched between different cores, potentially reducing the efficiency of memory access and caches.

The random \IO pattern shows the most significant performance loss when decreasing both RAM and CPU cores, with up to a \qty{60}{\percent} penalty in write operation with replica count \num{3}.
In particular, random write operations suffer notably from smaller amounts of RAM, while decreasing the core count yields a relatively small penalty (less than \qty{15}{\percent}).
Random read operations appear to be influenced by both CPU and RAM, with a clear relationship: when less RAM is available per OSD, increasing the core count has less impact.

Similar to sequential \IO operations, the CPU-limited behavior is observed during random read. 
As with sequential \IO, allocating \num{2.3} cores per OSD does not consistently yield the highest average performance and often results in high variance, especially during write operations. 

This confirms the observations made in sequential \IO, highlighting the need for further investigation, possibly involving proper process binding and core affinity setup since the default CPU mask for OSD processes ---used in the experiments--- allows them to run on all cores.

\paragraph{Impact of HDD speed}

One of the most interesting findings concerns the relationship between hard drive performance and \ceph  performance. The sequential read experiments show that slower devices achieved similar performance to faster devices. Sequential writes, on the other hand, benefit from faster drives—but only up to a certain threshold, with a performance loss between up to \qty{40}{\percent}. A performance penalty arises using hard drives slower than \qty{175}{\mega\byte\per\second}, a speed easily reached even by modern low-end hard drives.

A similar trend is observed with random I/O patterns. The first performance plateau occurs with devices rated at below \qty{175}{\mega\byte\per\second} —with a performance drop of up to \qty{50}{\percent}. A second, higher plateau is reached when using devices faster than \qty{175}{\mega\byte\per\second}.

We can speculate that the limitation on the drives is not solely due to bandwidth, but also to latency. This parameter plays a crucial role in the final system performance, making a lower latency preferable over a higher bandwidth beyond a certain threshold. Our experiments are performed by design at constant latency, since it is a mechanical characteristic of the devices and cannot be controlled by the software. 
Eventually, the \gls{cgroups} \emph{blockio} controller allows tuning the maximum read and write \iops per device, providing a proxy of the physical latency limited to a specific case. 

\section{Conclusion}

While the results presented here apply to the \cephfs filesystem, the methodology is general enough to be applied to other well-known parallel \fss—such as Lustre, BeeGFS, or GPFS—to gain insight into different workload profiles.

Our results for sequential \IO patterns indicate that high-end, powerful machines are not necessarily required to achieve strong performance. Therefore, long-term storage systems designed primarily for storing or retrieving large volumes of data sequentially can operate effectively with a limited amount of RAM and a modest core count.

In contrast, \fss that are accessed frequently with unpredictable \IO patterns benefit significantly from large amounts of RAM, and to some extent, higher CPU core counts.

Interestingly, we found that increasing hard drive speed beyond approximately \qty{200}{\mega\byte\per\second} does not yield noticeable performance gains; this suggest that investing in lower-latency drives could be more beneficial than aiming for higher bandwidth—even in the case of sequential \IO.

We emphasize that our experiments reflect typical operational conditions and do not cover exceptional scenarios such as node failure recovery or driver replacement. Such events may heavily stress the storage nodes and should be considered when provisioning resources, ensuring sufficient computing capacity.

Finally, we note that the configurations tested here use significantly fewer resources than those recommended in the official documentation \cite{cephHardware}. 

Future direction includes a more detailed exploration of data storage layouts; for instance, the use of erasure coding—known to be computationally intensive—could yield very different performance outcomes.

\begin{acks}
The authors acknowledge the AREA Science Park supercomputing platform ORFEO made available for conducting the research reported in this paper and the technical support of the Laboratory of Data Engineering staff. 

I.P. was supported by the Programma Nazionale della Ricerca (PNR) grant J95F21002830001 with the title “FAIR-by-design”. 
\end{acks}


\bibliographystyle{ACM-Reference-Format}
\bibliography{ourbib}

\newpage
\appendix

\section{\fio job files}

\label{sec:fio-job}
This section presents the global and per-job options used in the \fio job files. To ensure consistent and cache-free measurements, we explicitly enabled all relevant options to bypass client-side caches. This includes disabling buffered \IO, enabling direct \IO, and invalidating any existing page cache—even when some options may overlap in effect depending on the selected I/O engine.

\begin{lstlisting}[caption={Global \fio job file configuration used to benchmark \cephfs. This setup applies to experiments described in \autoref{subsed:memory_core_result} and \autoref{subsed:hdd_result}.}, label={lst:fio}, language=c]
[global]
directory=/mnt/benchmark/replica1/
ioengine=posixaio
direct=1
buffered=0
invalidate=1
time_based=1
runtime=150
size=16G
group_reporting=1
stonewall
\end{lstlisting}

\begin{lstlisting}[caption={\fio job configuration sections used for random and sequential read/write benchmarks.
Each section defines a different workload with specific block size and \IO pattern.}, label={lst:fio-detail}, language=c]
[Randread]
rw=randread
bs=4k
numjobs=64

[Read]
rw=read
bs=8m
numjobs=64

[Randwrite]
rw=randwrite
bs=4k
numjobs=64

[Write]
rw=write
bs=8m
numjobs=64
\end{lstlisting}

\section{Raw Performance Results}

In this section, we report the result in the non-normalized form. All the following tables report the average bandwidth measured in the different runs of the related experiments. 
In particular,  \autoref{tab:rw_details_sequential} and \autoref{tab:rw_details_random} refers to \autoref{subsed:memory_core_result}; instead \autoref{tab:disks_details_disks} is related to \autoref{subsed:hdd_result}.

\begin{table}[H]
  \centering
  \caption{Average bandwidth performance [\unit{\gibi\byte\per\second}] measured during sequential read (\autoref{subtab:rw_read}) and write (\autoref{subtab:rw_write}) operations to a \cephfs \filesystem under varying CPU/OSD and RAM/OSD configurations, comparing replica-\num{2} and replica-\num{3} pool configurations. Values are reported as mean $\pm$ standard deviation.}
  \label{tab:rw_details_sequential}

  \begin{subtable}{.45\textwidth}
    \centering
    \caption{Sequential  Read}
    \label{subtab:rw_read}
    \begin{tabular}{ccccc}
      \toprule
      \textbf{RAM/OSD} & 
      \makecell{\textbf{CPU/OSD}\\0.6} & 
      \makecell{\textbf{CPU/OSD}\\1.1} & 
      \makecell{\textbf{CPU/OSD}\\1.7} & 
      \makecell{\textbf{CPU/OSD}\\2.3} \\
      \midrule
      \multicolumn{5}{c}{\textbf{Replica 2}} \\
      \midrule
        \textbf{2.3 GB}  & 3.82 $\pm$ 0.55 & 3.87 $\pm$ 0.51 & 4.02 $\pm$ 0.70 & 5.96 $\pm$ 0.98 \\
        \textbf{4.6 GB}  & 3.43 $\pm$ 0.10 & 3.35 $\pm$ 0.18 & 3.46 $\pm$ 0.11 & 3.95 $\pm$ 0.74 \\
        \textbf{9.1 GB}  & 3.29 $\pm$ 0.20 & 3.44 $\pm$ 0.18 & 3.80 $\pm$ 0.52 & 4.18 $\pm$ 0.87 \\
        \textbf{13.7 GB} & 3.38 $\pm$ 0.33 & 4.90 $\pm$ 0.66 & 5.25 $\pm$ 0.04 & 4.47 $\pm$ 0.88 \\
      \midrule
      \multicolumn{5}{c}{\textbf{Replica 3}} \\
      \midrule
        \textbf{2.3 GB}  &  3.72 $\pm$ 0.56 & 4.12 $\pm$ 0.27 & 3.95 $\pm$ 0.81 & 6.00 $\pm$ 0.60 \\
        \textbf{4.6 GB}  & 3.39 $\pm$ 0.20 & 3.29 $\pm$ 0.24 & 3.41 $\pm$ 0.18 & 4.11 $\pm$ 1.02 \\
        \textbf{9.1 GB}  & 3.31 $\pm$ 0.29 & 3.35 $\pm$ 0.42 & 3.76 $\pm$ 0.46 & 3.94 $\pm$ 0.73 \\
        \textbf{13.7 GB} & 3.46 $\pm$ 0.15 & 4.69 $\pm$ 0.72 & 5.05 $\pm$ 0.03 & 4.58 $\pm$ 0.56 \\
      \bottomrule
    \end{tabular}
  \end{subtable}

  \vspace{0.5em}

  \begin{subtable}{.45\textwidth}
    \centering
    \caption{Sequential Write}
    \label{subtab:rw_write}
    \begin{tabular}{ccccc}
      \toprule
      \textbf{RAM/OSD} & 
      \makecell{\textbf{CPU/OSD}\\0.6} & 
      \makecell{\textbf{CPU/OSD}\\1.1} & 
      \makecell{\textbf{CPU/OSD}\\1.7} & 
      \makecell{\textbf{CPU/OSD}\\2.3} \\
      \midrule
      \multicolumn{5}{c}{\textbf{Replica 2}} \\
      \midrule
        \textbf{2.3 GB}  & 7.36 $\pm$ 0.69 & 7.85 $\pm$ 0.27 & 7.78 $\pm$ 0.35 & 7.76 $\pm$ 0.64 \\
        \textbf{4.6 GB}  & 7.07 $\pm$ 0.46 & 7.86 $\pm$ 0.26 & 7.84 $\pm$ 0.44 & 8.05 $\pm$ 0.22 \\
        \textbf{9.1 GB}  &  7.95 $\pm$ 1.15 & 8.72 $\pm$ 0.23 & 8.83 $\pm$ 0.37 & 8.55 $\pm$ 0.40 \\
        \textbf{13.7 GB} & 8.02 $\pm$ 0.33 & 8.66 $\pm$ 0.14 & 8.70 $\pm$ 0.15 & 8.38 $\pm$ 0.67 \\
      \midrule
      \multicolumn{5}{c}{\textbf{Replica 3}} \\
      \midrule
        \textbf{2.3 GB}  & 4.71 $\pm$ 0.56 & 5.42 $\pm$ 0.19 & 5.27 $\pm$ 0.33 & 5.16 $\pm$ 0.46 \\
        \textbf{4.6 GB}  &  4.99 $\pm$ 0.22 & 5.44 $\pm$ 0.21 & 5.59 $\pm$ 0.18 & 5.55 $\pm$ 0.30 \\
        \textbf{9.1 GB}  & 5.32 $\pm$ 0.73 & 5.75 $\pm$ 0.35 & 5.87 $\pm$ 0.09 & 5.70 $\pm$ 0.23 \\
        \textbf{13.7 GB} & 5.50 $\pm$ 0.28 & 5.89 $\pm$ 0.14 & 5.87 $\pm$ 0.25 & 5.70 $\pm$ 0.48 \\
      \bottomrule
    \end{tabular}
  \end{subtable}
\end{table}

\begin{table}[H]
  \centering
  \caption{Average bandwidth performance [$\iops \cdot 10^3$] measured during random read (\autoref{subtab:rndrw_read}) and write (\autoref{subtab:rndrw_write}) operations to a \cephfs \filesystem under varying CPU/OSD and RAM/OSD configurations, comparing replica-\num{2} and replica-\num{3} pool configurations. Values are reported as mean $\pm$ standard deviation.}
  \label{tab:rw_details_random}

  \begin{subtable}{.45\textwidth}
    \centering
    \caption{Random Read}
    \label{subtab:rndrw_read}
    \begin{tabular}{ccccc}
      \toprule
      \textbf{RAM/OSD} & 
      \makecell{\textbf{CPU/OSD}\\0.6} & 
      \makecell{\textbf{CPU/OSD}\\1.1} & 
      \makecell{\textbf{CPU/OSD}\\1.7} & 
      \makecell{\textbf{CPU/OSD}\\2.3} \\
      \midrule
      \multicolumn{5}{c}{\textbf{Replica 2}} \\
      \midrule
        \textbf{2.3 GB}  & 13.5 $\pm$ 1.1 & 14.0 $\pm$ 0.7 & 14.9 $\pm$ 1.5 & 15.5 $\pm$ 1.4 \\
        \textbf{4.6 GB}  & 13.2 $\pm$ 1.5 & 12.5 $\pm$ 0.7 & 12.6 $\pm$ 0.6 & 13.6 $\pm$ 1.3 \\
        \textbf{9.1 GB}  & 12.3 $\pm$ 1.7 & 14.8 $\pm$ 1.4 & 16.2 $\pm$ 2.2 & 17.5 $\pm$ 2.0 \\
        \textbf{13.7 GB} & 12.0 $\pm$ 0.9 & 15.8 $\pm$ 4.0 & 19.8 $\pm$ 0.2 & 19.3 $\pm$ 2.3 \\
      \midrule
      \multicolumn{5}{c}{\textbf{Replica 3}} \\
      \midrule
        \textbf{2.3 GB}  & 12.1 $\pm$ 0.5 & 12.6 $\pm$ 0.3 & 12.8 $\pm$ 0.6 & 13.2 $\pm$ 0.9 \\
        \textbf{4.6 GB}  & 13.0 $\pm$ 0.6 & 12.8 $\pm$ 0.3 & 12.6 $\pm$ 0.6 & 13.4 $\pm$ 0.4 \\
        \textbf{9.1 GB}  & 13.6 $\pm$ 0.8 & 15.5 $\pm$ 1.6 & 17.1 $\pm$ 1.8 & 17.8 $\pm$ 1.1 \\
        \textbf{13.7 GB} & 13.5 $\pm$ 0.4 & 15.8 $\pm$ 4.9 & 20.5 $\pm$ 0.1 & 19.8 $\pm$ 2.0 \\
      \bottomrule
    \end{tabular}
  \end{subtable}

  \vspace{0.5em}

  \begin{subtable}{.45\textwidth}
    \centering
    \caption{Random Write}
    \label{subtab:rndrw_write}
    \begin{tabular}{ccccc}
      \toprule
      \textbf{RAM/OSD} & 
      \makecell{\textbf{CPU/OSD}\\0.6} & 
      \makecell{\textbf{CPU/OSD}\\1.1} & 
      \makecell{\textbf{CPU/OSD}\\1.7} & 
      \makecell{\textbf{CPU/OSD}\\2.3} \\
      \midrule
      \multicolumn{5}{c}{\textbf{Replica 2}} \\
      \midrule
        \textbf{2.3 GB}  & 10.3 $\pm$ 0.6 & 10.2 $\pm$ 0.5 & 10.7 $\pm$ 0.6 & 10.3 $\pm$ 1.0 \\
        \textbf{4.6 GB}  & 16.4 $\pm$ 0.6 & 15.2 $\pm$ 1.3 & 16.0 $\pm$ 0.6 & 15.4 $\pm$ 1.0 \\
        \textbf{9.1 GB}  & 21.0 $\pm$ 0.5 & 21.6 $\pm$ 0.6 & 22.3 $\pm$ 0.9 & 20.8 $\pm$ 1.8 \\
        \textbf{13.7 GB} & 21.4 $\pm$ 0.7 & 21.7 $\pm$ 1.0 & 23.6 $\pm$ 0.5 & 22.9 $\pm$ 3.2 \\
      \midrule
      \multicolumn{5}{c}{\textbf{Replica 3}} \\
      \midrule
        \textbf{2.3 GB}  & 6.6 $\pm$ 0.3 & 6.4 $\pm$ 0.4 & 6.2 $\pm$ 0.5 & 6.1 $\pm$ 1.3 \\
        \textbf{4.6 GB}  & 10.4 $\pm$ 0.7 & 9.8 $\pm$ 0.7 & 10.0 $\pm$ 0.3 & 10.0 $\pm$ 0.9 \\
        \textbf{9.1 GB}  &  15.5 $\pm$ 0.4 & 14.8 $\pm$ 0.6 & 15.4 $\pm$ 0.4 & 14.6 $\pm$ 0.9 \\
        \textbf{13.7 GB} & 15.1 $\pm$ 0.5 & 15.7 $\pm$ 0.3 & 16.2 $\pm$ 0.2 & 15.6 $\pm$ 2.8 \\
      \bottomrule
    \end{tabular}
  \end{subtable}
\end{table}

\begin{table}[H]
  \centering
  \caption{Average bandwidth performance measured during sequential (\autoref{subtab:disk_det_seq_rep2} and \autoref{subtab:disk_det_seq_rep3}) and random (\autoref{subtab:disk_det_rand_rep2} and \autoref{subtab:disk_det_rand_rep3}) operations to a \cephfs \filesystem under varying device speeds, comparing replica-\num{2} and replica-\num{3} pool configurations. Values are reported as mean $\pm$ standard deviation.
  }
  \label{tab:disks_details_disks}
  
  \begin{subtable}[t]{0.48\textwidth}
    \centering
    \caption{Sequential Performance [\unit{\gibi\byte\per\second}] – Replica-\num{2}}
    \label{subtab:disk_det_seq_rep2}
    \begin{tabular}{ccc}
      \toprule
      Device speed [MB/s] & Read & Write \\
      \midrule
          100 & 2.20 $\pm$ 0.29 & 2.12 $\pm$ 0.10 \\
          125 & 1.95 $\pm$ 0.60 & 2.65 $\pm$ 0.03 \\
          150 & 2.21 $\pm$ 0.44 & 2.97 $\pm$ 0.17 \\
          175 & 2.10 $\pm$ 0.59 & 3.25 $\pm$ 0.43 \\
          200 & 1.93 $\pm$ 0.61 & 3.50 $\pm$ 0.29 \\
          225 & 1.99 $\pm$ 0.66 & 3.55 $\pm$ 0.09 \\
          250 & 2.30 $\pm$ 0.48 & 3.42 $\pm$ 0.12 \\
          275 & 2.27 $\pm$ 0.45 & 3.40 $\pm$ 0.16 \\
      \bottomrule
    \end{tabular}
  \end{subtable}
  \hfill
  \begin{subtable}[t]{0.48\textwidth}
    \centering
    \caption{Sequential Performance [\unit{\gibi\byte\per\second}] – Replica-\num{3}}
    \label{subtab:disk_det_seq_rep3}
    \begin{tabular}{ccc}
      \toprule
      Device speed [MB/s] & Read & Write \\
      \midrule
          100 & 2.30 $\pm$ 0.04 & 1.44 $\pm$ 0.08 \\
          125 & 2.23 $\pm$ 0.24 & 1.80 $\pm$ 0.08 \\
          150 & 2.36 $\pm$ 0.06 & 2.08 $\pm$ 0.18 \\
          175 & 2.16 $\pm$ 0.35 & 2.33 $\pm$ 0.09 \\
          200 & 1.99 $\pm$ 0.46 & 2.34 $\pm$ 0.15 \\
          225 & 2.09 $\pm$ 0.32 & 2.25 $\pm$ 0.11 \\
          250 & 2.27 $\pm$ 0.30 & 2.35 $\pm$ 0.09 \\
          275 & 2.21 $\pm$ 0.42 & 2.27 $\pm$ 0.09 \\
      \bottomrule
    \end{tabular}
  \end{subtable}

  \vspace{0.5em}

  \begin{subtable}[t]{0.48\textwidth}
    \centering
    \caption{Random Performance [IOPS $\cdot 10^3$] – Replica-2}
    \label{subtab:disk_det_rand_rep2}
    \begin{tabular}{ccc}
      \toprule
      Device speed [MB/s] & Read & Write \\
      \midrule
         100 & 4.62 $\pm$ 0.07 & 10.05 $\pm$ 0.50 \\
          125 & 4.55 $\pm$ 0.15 & 10.13 $\pm$ 0.68 \\
          150 & 4.76 $\pm$ 0.05 & 8.93 $\pm$ 0.90 \\
          175 & 7.72 $\pm$ 2.09 & 8.24 $\pm$ 0.58 \\
          200 & 8.85 $\pm$ 0.42 & 8.34 $\pm$ 0.98 \\
          225 & 8.94 $\pm$ 0.35 & 8.67 $\pm$ 0.46 \\
          250 & 8.75 $\pm$ 0.37 & 8.30 $\pm$ 0.82 \\
          275 & 8.87 $\pm$ 0.41 & 8.48 $\pm$ 0.76 \\
      \bottomrule
    \end{tabular}
  \end{subtable}
  \hfill
  \begin{subtable}[t]{0.48\textwidth}
    \centering
    \caption{Random Performance [IOPS $\cdot 10^3$] – Replica-3}
    \label{subtab:disk_det_rand_rep3}
    \begin{tabular}{ccc}
      \toprule
      Device speed [MB/s] & Read & Write \\
      \midrule
          100 & 4.84 $\pm$ 0.10 & 7.03 $\pm$ 0.54 \\
          125 & 4.76 $\pm$ 0.20 & 7.03 $\pm$ 0.37 \\
          150 & 4.72 $\pm$ 0.04 & 6.46 $\pm$ 0.51 \\
          175 & 6.66 $\pm$ 1.37 & 5.16 $\pm$ 0.27 \\
          200 & 7.74 $\pm$ 0.26 & 5.55 $\pm$ 0.38 \\
          225 & 8.06 $\pm$ 0.10 & 6.13 $\pm$ 0.39 \\
          250 & 7.91 $\pm$ 0.23 & 5.60 $\pm$ 0.58 \\
          275 & 8.03 $\pm$ 0.07 & 5.92 $\pm$ 0.26 \\
      \bottomrule
    \end{tabular}
  \end{subtable}
\end{table}

\end{document}